\documentclass{article}
\pdfoutput=1
\usepackage{arxiv}
\usepackage[utf8]{inputenc} 
\usepackage[T1]{fontenc}    
\usepackage{amsmath,amssymb,amsthm}
\usepackage{mathtools}
\usepackage{enumitem}
\usepackage{booktabs}
\usepackage{microtype}      
\usepackage{doi}
\usepackage{array}
\usepackage{hyperref}
\usepackage{cleveref}
\usepackage{algorithm}
\usepackage{algpseudocode}
\usepackage[most]{tcolorbox}
\usepackage{tikz}
\usetikzlibrary{
	arrows,	arrows.meta,backgrounds,bending,calc,decorations.markings,decorations.pathmorphing,decorations.pathreplacing,fit,intersections,positioning,scopes,shapes,shapes.geometric,shapes.multipart,tikzmark,3d}
\tikzset{middlearrow/.style={
		decoration={markings,
			mark= at position 0.5 with {\arrow{#1}} ,
		},
		postaction={decorate}
	}
}

\tikzset{
	partial ellipse/.style args={#1:#2:#3}{
		insert path={+ (#1:#3) arc (#1:#2:#3)}
	}
}

\colorlet{color 0}{white}
\colorlet{color 1}{black}

\tikzset{
	random dot/.style args={#1*#2:#3}{random dot*={#1}*{#2}:{#3} and {#3}},
	random dot*/.style args={#1*#2:#3 and #4}{
		insert path={\foreach \throwaway in {1,...,#1} {(canvas polar cs: angle={#2+rnd}, x radius={#3+#3/.5cm*rand}, y radius={{#4+#4/.5cm*rand}}) circle[]}}}
}

\usepackage{mathtools}

\definecolor{boxcolor1}{RGB}{255,254,230}	
\definecolor{boxcolor2}{RGB}{252,227,215}	
\definecolor{boxcolor3}{RGB}{225,240,227}	
\definecolor{boxcolor4}{RGB}{227,235,246}	

\newtcolorbox{definitionbox}[1][]{
	breakable,
	before skip=1.5\topskip,
	after skip=1.5\topskip,
	left skip=0pt,
	right skip=0pt,
	left=4pt,
	right=4pt,
	top=2pt,
	bottom=2pt,
	lefttitle=4pt,
	righttitle=4pt,
	toptitle=2pt,
	bottomtitle=2pt,
	sharp corners,
	boxrule=0pt,
	titlerule=.4pt,
	colback=boxcolor1,
	colbacktitle=boxcolor1,
	coltitle=black,
	colframe=darkgray,
	coltext=black,
	fonttitle=\bfseries,
	title=Definition~\thetcbcounter:,
	#1
}

\newtcolorbox{theorembox}[1][]{
	breakable,
	before skip=1.5\topskip,
	after skip=1.5\topskip,
	left skip=0pt,
	right skip=0pt,
	left=4pt,
	right=4pt,
	top=2pt,
	bottom=2pt,
	lefttitle=4pt,
	righttitle=4pt,
	toptitle=2pt,
	bottomtitle=2pt,
	sharp corners,
	boxrule=0pt,
	titlerule=.4pt,
	colback=boxcolor2,
	colbacktitle=boxcolor2,
	coltitle=black,
	colframe=darkgray,
	coltext=black,
	fonttitle=\bfseries,
	title=Theorem~\thetcbcounter:,
	#1
}

\newtcolorbox{lemmabox}[1][]{
	breakable,
	before skip=1.5\topskip,
	after skip=1.5\topskip,
	left skip=0pt,
	right skip=0pt,
	left=4pt,
	right=4pt,
	top=2pt,
	bottom=2pt,
	lefttitle=4pt,
	righttitle=4pt,
	toptitle=2pt,
	bottomtitle=2pt,
	sharp corners,
	boxrule=0pt,
	titlerule=.4pt,
	colback=boxcolor3,
	colbacktitle=boxcolor3,
	coltitle=black,
	colframe=darkgray,
	coltext=black,
	fonttitle=\bfseries,
	title=Lemma~\thetcbcounter:,
	#1
}

\newtcolorbox{propositionbox}[1][]{
	breakable,
	before skip=1.5\topskip,
	after skip=1.5\topskip,
	left skip=0pt,
	right skip=0pt,
	left=4pt,
	right=4pt,
	top=2pt,
	bottom=2pt,
	lefttitle=4pt,
	righttitle=4pt,
	toptitle=2pt,
	bottomtitle=2pt,
	sharp corners,
	boxrule=0pt,
	titlerule=.4pt,
	colback=boxcolor3,
	colbacktitle=boxcolor3,
	coltitle=black,
	colframe=darkgray,
	coltext=black,
	fonttitle=\bfseries,
	title=Proposition~\thetcbcounter:,
	#1
}

\newtcolorbox{corollarybox}[1][]{
	breakable,
	before skip=1.5\topskip,
	after skip=1.5\topskip,
	left skip=0pt,
	right skip=0pt,
	left=4pt,
	right=4pt,
	top=2pt,
	bottom=2pt,
	lefttitle=4pt,
	righttitle=4pt,
	toptitle=2pt,
	bottomtitle=2pt,
	sharp corners,
	boxrule=0pt,
	titlerule=.4pt,
	colback=boxcolor4,
	colbacktitle=boxcolor4,
	coltitle=black,
	colframe=darkgray,
	coltext=black,
	fonttitle=\bfseries,
	title=Corollary~\thetcbcounter:,
	#1
}

\newtheoremstyle{mystyle}
{3pt}
{3pt}
{\itshape}
{}
{\bfseries}
{.}
{.5em}
{}

\theoremstyle{mystyle}

\newtheorem{definition}{Definition}[section] 

\renewenvironment{definition}{%
	\refstepcounter{definition}
	\begin{definitionbox}[title=Definition~\thedefinition]
	}{%
	\end{definitionbox}
}

\newtheorem{theorem}{Theorem}[section] 

\renewenvironment{theorem}{%
	\refstepcounter{theorem}
	\begin{theorembox}[title=Theorem~\thetheorem]
	}{%
	\end{theorembox}
}

\newtheorem{lemma}{Lemma}[section] 

\renewenvironment{lemma}{%
	\refstepcounter{lemma}
	\begin{lemmabox}[title=Lemma~\thelemma]
	}{%
	\end{lemmabox}
}

\newtheorem{proposition}{Proposition}[section] 

\renewenvironment{proposition}{%
	\refstepcounter{proposition}
	\begin{propositionbox}[title=Proposition~\theproposition]
	}{%
	\end{propositionbox}
}

\newtheorem{corollary}{Corollary}[section] 

\renewenvironment{corollary}{%
	\refstepcounter{corollary}
	\begin{corollarybox}[title=Corollary~\thecorollary]
	}{%
	\end{corollarybox}
}

\newtheoremstyle{claimstyle} 
{3pt} 
{3pt} 
{} 
{} 
{\bfseries} 
{.} 
{0.5em} 
{} 

\theoremstyle{claimstyle}

\newtheoremstyle{postulatestyle} 
{3pt} 
{3pt} 
{} 
{} 
{\bfseries} 
{.} 
{0.5em} 
{} 

\theoremstyle{postulatestyle}

\newtheoremstyle{questionstyle} 
{3pt} 
{3pt} 
{} 
{} 
{\bfseries} 
{.} 
{0.5em} 
{} 

\theoremstyle{questionstyle}

\newtheoremstyle{assumptionstyle} 
{3pt} 
{3pt} 
{} 
{} 
{\bfseries} 
{.} 
{0.5em} 
{} 

\theoremstyle{assumptionstyle}

\newtheoremstyle{conjecturestyle} 
{3pt} 
{3pt} 
{} 
{} 
{\bfseries} 
{.} 
{0.5em} 
{} 

\theoremstyle{conjecturestyle}

\newtheoremstyle{critiquestyle} 
{3pt} 
{3pt} 
{\itshape} 
{} 
{\bfseries} 
{.} 
{0.5em} 
{} 

\theoremstyle{critiquestyle}

\newtheoremstyle{responsestyle} 
{3pt} 
{3pt} 
{\itshape} 
{} 
{\bfseries} 
{.} 
{0.5em} 
{} 

\theoremstyle{responsestyle}
\title{Mean tropical year length at arbitrary ecliptic longitude}

\author{ \href{https://orcid.org/0009-0004-7957-1806}{\includegraphics[scale=0.06]{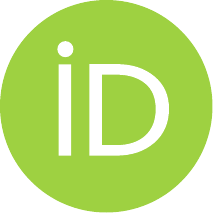}\hspace{1mm}Daniel Quigley} \\
	Center for Possible Minds\\
	Indiana University Bloomington\\
	Bloomington, IN 47408 \\
	\texttt{dgquigle@iu.edu} \\
}

\renewcommand{\headeright}{Mean year lengths}
\renewcommand{\undertitle}{Preprint}
\renewcommand{\shorttitle}{Mean year lengths}
\colorlet{linecol}{cyan!90!blue!90!black}
\colorlet{fillcol}{cyan!60!blue!80!black!40}
\hypersetup{
	pdftitle={Mean tropical year length at arbitrary ecliptic longitude},
	pdfsubject={solar year calculation, derivation of mean year length},
	pdfauthor={Daniel Quigley},
	pdfkeywords={tropical year, ecliptic longitude, solar longitude, orbital eccentricity, apsidal precession},
}

\begin{document}
	\maketitle
	
	\begin{abstract}
		We compute the mean interval between successive returns of the apparent geocentric solar longitude $\lambda$ to a fixed value $L \in \{0^\circ, 45^\circ, 90^\circ, \ldots, 315^\circ\}$, averaged over a multi-millennium window; this gives eight ``mean years'' against which calendar leap rules can be tuned: four cardinal-point years (equinoxes and solstices); four cross-quarter years. The construction is built on Meeus's low-precision solar theory (\textit{Astronomical Algorithms}, 2nd ed., 1998), itself a low-order truncation of Newcomb's \textit{Tables of the Sun} re-expanded around J2000.0. Where Meeus presents polynomial coefficients without justification, we draw on Smart's \textit{Textbook on Spherical Astronomy} (6th ed., revised by Green, 1977) for the underlying derivations. Numerical accuracy is validated against the cardinal-point intervals tabulated in Meeus, \textit{More Mathematical Morsels}, 2002. We close with a derivation of the secular drift equation showing that, regardless of how well a leap rule is tuned, the slow shrinkage of the tropical year produces a quadratic cumulative error that reaches one day in $\sim$5{,}700 years for any fixed intercalation rule.

	\end{abstract}

	
	\section{Introduction}
	
	The mean year at ecliptic longitude $L$, denoted $Y(L)$, is the mean interval between successive returns of the apparent geocentric solar longitude to the value $L \in [0^\circ, 360^\circ)$, averaged over a multi-century window. For $L = 0^\circ$ (the March equinox), $Y(0^\circ)$ is the familiar vernal-equinox tropical year; for $L = 90^\circ$, $180^\circ$, $270^\circ$ it reduces to the June-solstice, September-equinox, and December-solstice tropical years tabulated by \cite{MeeusSavoie1992} and \cite{Meeus2002}. Treating $L$ as a continuous variable rather than sampling it only at the four cardinal points yields a smooth near-sinusoidal curve, spanning $\sim98$ seconds peak-to-peak at the present epoch, whose shape encodes the interaction of orbital eccentricity with the slow drift of the Earth's perihelion relative to the equinoxes.
	
	We derive $Y(L)$ and provide a self-contained computational procedure for reproducing and extending the results. Sections~\ref{sec:timeargument}--\ref{sec:meanlongitude} establish the time argument and the mean-longitude polynomial. Section~\ref{sec:precession} disentangles the three rates encoded in the Meeus polynomial coefficients, and derives the general precession in longitude following \cite{Smart1977} and \cite{BrouwerClemence1961}: tropical mean motion; sidereal mean motion; anomalistic mean motion. Section~\ref{sec:eqncenter} derives the equation of center from the series inversion of Kepler's equation, connecting Meeus's numerical coefficients directly to the Earth's eccentricity $e$. Section~\ref{sec:abberrationnutation} accounts for aberration and the leading nutation term. Section~\ref{sec:rootfinding} describes the modified Newton iteration used to locate individual crossings of longitude $L$. Section~\ref{sec:meanerror} constructs the mean interval $\bar{Y}_L$ from a sequence of crossings and establishes its error, in which we show that $\Delta T$ does not enter $\bar{Y}_L$ at all, since it is a TT interval, and that the formula's $\sim 0.01^\circ$ instantaneous error does not materially affect $\bar{Y}_L$ when $N \sim 3000$ intervals are averaged. Section~\ref{sec:j2000b1900} justifies the choice of the J2000 polynomial expansion over the older B1900 form. Section~\ref{sec:meanyears} assembles the procedure to recover eight representative $Y(L)$ values, validates against \cite{Meeus2002}, presents the full $Y(L)$ curve at $1^\circ$ resolution with a physical explanation of its structure, tabulates the longitude-averaged $\langle Y\rangle$ against published reference values with a discussion of the distinction between the mean tropical and vernal-equinox years, and quantifies the epoch dependence of all eight marker year lengths across seven millennia. Section~\ref{sec:secular} closes with some remarks on secular drift and its consequences for imposing a fixed year length and finite leap rules.
	
	The derivation follows \cite{Meeus1998} for the solar-longitude polynomial, \cite{Smart1977} for the underlying spherical-astronomy derivations, and \cite{BrouwerClemence1961} for the celestial-mechanics on precession and the series inversion of Kepler's equation.  Numerical results are validated against the cardinal-point year lengths tabulated in \cite{MeeusSavoie1992} and \cite{Meeus2002}; the maximum residual at the four cardinal points is $2.2$~s, consistent with the error bound derived in Section~\ref{sec:meanerror}.  All computations are implemented in open-source software, available at the author's GitHub.
	
	Two terminological notes. First, the labels ``Beltane'', ``Lughnasadh,'' ``Samhain'', and ``Imbolc'' appear in Table~\ref{tab:values} as shorthand for the ecliptic longitudes $L = 45^\circ$, $135^\circ$, $225^\circ$, and $315^\circ$, respectively; their use here is purely geometric: these are the longitude midpoints between the cardinal events. The association of these Celtic festival names with specific ecliptic longitudes is a twentieth-century analytical construct \cite{Hutton2001,McCluskey1989}; no historical claim is intended. Second, we work throughout in Terrestrial Time (TT); Section~\ref{sec:meanerror} shows this is harmless at the precision level of the computation.
	
	\section{Time argument and epoch}\label{sec:timeargument}
	
	Let JD denote Julian Date (TT scale; we ignore $\Delta T$; see Section~\ref{sec:meanerror} for why this is harmless). Define Julian centuries from J2000.0:
	\[
	T = \frac{\mathrm{JD} - 2451545.0}{36525}
	\]
	The epoch JD~2451545.0 is 2000 Jan~1.5 TT. This is the reference time around which the polynomial coefficients throughout are expanded; Section~\ref{sec:j2000b1900} justifies this explicitly against the older B1900 expansion still found in \cite{Meeus1988, Meeus1998}.
	
	\section{Mean longitude and mean anomaly}\label{sec:meanlongitude}
	
	Geometric mean longitude of the Sun, referred to the mean equinox of the date \cite{Meeus1998}:
	\[
	L_0(T) = 280.46646^\circ + 36000.76983^\circ\,T + 0.0003032^\circ\,T^2
	\]
	Mean anomaly \cite{Meeus1998}:
	\[
	M(T) = 357.52911^\circ + 35999.05029^\circ\,T - 0.0001537^\circ\,T^2
	\]
	The phrase ``referred to the mean equinox of the date'' does much heavy lifting here: the equinox itself is moving in inertial space, and the linear coefficient of $36000.76983^\circ$/century is \emph{not} the Sun's geometric mean motion against the stars, but the geometric mean motion \emph{plus the rate at which the equinox slides westward}. We disentangle the two in Section~\ref{sec:precession}.
	
	The linear coefficient of $L_0$, divided by $36525$ days, yields
	\[
	\dot\lambda_\mathrm{mean} = 0.9856473^\circ/\mathrm{day},
	\]
	the constant slope used by the Newton iterator in Section~\ref{sec:rootfinding}. The fact that $L_0$ and $M$ have different linear coefficients that differ by $1.7195^\circ$/century is itself a physical statement, which we address below in Section~\ref{sec:precession}.
	
	\section{Precession of the equinoxes}\label{sec:precession}
	
	The \textbf{equinox} is the line where Earth's equatorial plane intersects the ecliptic plane. Both planes move slowly in inertial space, but the equatorial plane moves much faster, and the line of intersection (the equinox direction) accordingly precesses westward along the ecliptic at $\sim50.29''$/yr, completing a full circuit in $\sim25770$ years. This is general precession in longitude.
	
	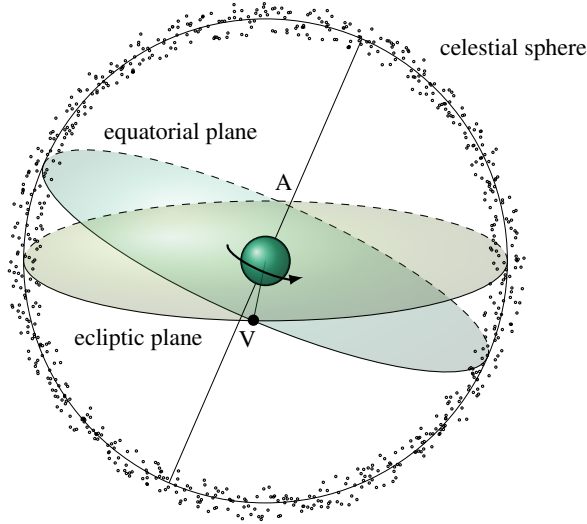
\begin{figure}[H]
		\centering
		\begin{tikzpicture}[style={radius=.4pt+.2pt*rand},scale=0.80]
			
			
			\node[] at (4.1,3.5) {\small{celestial sphere}};
			\node[] at (-1.4,2.1) {\small{equatorial plane}};
			\node[] at (-2.1,-1.3) {\small{ecliptic plane}};
			\node[] at (-0.3,-1.3) {\small{V}};
			\node[] at (0.3,1.3) {\small{A}};
			
			\draw [ultra thin] (0.0,0.0) -- (-0.2,-0.98);
			
			\draw[rotate=66.5,dashed,rotate=-90] (4,0) arc(0:180:4cm and 0.99cm);
			\draw[rotate=66.5,rotate=-90] (4,0) arc(360:180:4cm and 0.99cm);
			\shade[ball color=green!60!blue,opacity=0.20,rotate=-23.5] (0,0) ellipse (4cm and 0.99cm);
			
			\draw[rotate=89.94,dashed,rotate=-90] (4,0) arc(0:180:4cm and 0.99cm);
			\draw[rotate=89.94,rotate=-90] (4,0) arc(360:180:4cm and 0.99cm);
			\shade[ball color=green!60!red,opacity=0.20] (0,0) ellipse (4cm and 0.99cm);
			
			\node at (-0.2,-0.98)[circle,fill,inner sep=1.5pt]{};
			\draw [ultra thin,rotate=-23.5] (0,4) -- (0,0.4);
			\draw [ultra thin,rotate=-23.5] (0,-4) -- (0,-0.4);
			
			\draw[thick] (0,0) circle (0.4cm);
			\shade[ball color=green!60!blue,opacity=0.80] (0,0) circle (0.4cm);
			
			\draw[thick, black,<-,>=latex,rotate=-23.5] (0,0) [partial ellipse=360:180:0.68cm and 0.18cm];
			
			\foreach \angle in {0,...,360}
			\draw[black] [random dot=2*\angle:4cm];
			\draw[ultra thin] (4,0) arc(0:360:4cm and 4cm);
			
		\end{tikzpicture}
		\caption{Vernal equinox at point V, as the primary direction, and the autumnal equinox at point A; geographic poles projected onto the celestial sphere as celestial poles.}\label{fig:pol}
	\end{figure}

	\subsection{Physical causes of precession}\label{subsec:precession}
	
	Let us elaborate on the physical cause \cite{Smart1977}. Contrary to convenient depiction, the Earth is not spherical; its equatorial radius exceeds its polar radius by $\sim21$ km, an oblateness produced by rotation. Because the Earth's rotation axis is tilted by $\epsilon \approx 23.4^\circ$ to the ecliptic, the Sun and the Moon each spend half of their cycle north of the equatorial plane and half south of it; the pull on the near and far sides of the bulge is unequal, and the orbit-averaged effect is a net torque whose direction is perpendicular to both the rotation axis and the line to the perturber. The Moon dominates, contributing roughly $2.2\times$ the Sun's torque despite its far smaller mass, by virtue of its proximity (the torque scales as $m/r^3$) and the inclination of its orbit.
	
	For a perturber of mass $m$ at distance $r$ (treated as a point mass and orbit-averaged; we are reminded of a spherical cow) the torque on the Earth (a symmetric top with polar/equatorial moments $C > A$) is
	\[
	\bar\tau = -\frac{3Gm}{2r^3}(C - A)\sin\epsilon\cos\epsilon
	\]
	directed along the line of nodes. The Earth's spin angular momentum $L_z = C\omega$ responds gyroscopically: instead of tipping over, the axis precesses about the ecliptic pole. The axis of rotation traces out a circle about which the equatorial poles follow. The precession rate is
	\[
	\dot\psi = \frac{|\bar\tau|}{L_z\sin\epsilon} = \frac{3Gm}{2\omega r^3}\cdot\frac{C - A}{C}\cos\epsilon
	\]
	
	\begin{figure}[H]
		\centering
		\begin{tikzpicture}[style={radius=.4pt+.2pt*rand},scale=0.80]
			
			
			\draw [ultra thin] (0,3.6) -- (0,0.4);
			\draw [ultra thin] (0,-3.6) -- (0,-0.4);
			\node at (0,3.65)[circle,fill,inner sep=1.0pt]{};
			\draw[thick,black] (0,3.65) [partial ellipse=360:180:1.59cm and 0.18cm];
			\draw[thick,black,->,>=latex] (0,3.65) [partial ellipse=360:290:1.59cm and 0.18cm];
			\draw[thick,black,dashed] (0,3.65) [partial ellipse=0:180:1.59cm and 0.18cm];
			
			\node at (0,-3.65)[circle,fill,inner sep=1.0pt]{};
			\draw[thick,black] (0,-3.65) [partial ellipse=360:180:1.59cm and 0.18cm];
			\draw[thick,black,->,>=latex] (0,-3.65) [partial ellipse=360:240:1.59cm and 0.18cm];
			\draw[thick,black,dashed] (0,-3.65) [partial ellipse=0:180:1.59cm and 0.18cm];
			
			\node[] at (4.1,3.5) {\small{celestial sphere}};
			\node[] at (-1.4,2.1) {\small{equatorial plane}};
			\node[] at (-2.1,-1.3) {\small{ecliptic plane}};
			\node[] at (1.9,4.2) {\small{P\textsubscript{N}}};
			\node[] at (-1.9,-4.2) {\small{P\textsubscript{N}}};
			\node[] at (-0.3,-1.3) {\small{V}};
			\node[] at (0.3,1.3) {\small{A}};
			
			\draw [ultra thin] (0,0) -- (-0.2,-0.98);
			\draw [ultra thin] (0.2,0.98) -- (0.0799524,0.391928);
			
			\draw[rotate=66.5,dashed,rotate=-90] (4,0) arc(0:180:4cm and 0.99cm);
			\draw[rotate=66.5,rotate=-90] (4,0) arc(360:180:4cm and 0.99cm);
			\shade[ball color=green!60!blue,opacity=0.20,rotate=-23.5] (0,0) ellipse (4cm and 0.99cm);
			
			\draw[rotate=89.94,dashed,rotate=-90] (4,0) arc(0:180:4cm and 0.99cm);
			\draw[rotate=89.94,rotate=-90] (4,0) arc(360:180:4cm and 0.99cm);
			\shade[ball color=green!60!red,opacity=0.20] (0,0) ellipse (4cm and 0.99cm);
			
			\draw[very thick,rotate=-1.0,->,>=latex] (-0.25,-1) arc(268:240:4cm and 0.99cm);
			\draw[very thick,rotate=179.1,->,>=latex] (-0.25,-1) arc(268:240:4cm and 0.99cm);
			
			\node at (-0.2,-0.98)[circle,fill,inner sep=1.5pt]{};
			\node at (0.2,0.98)[circle,fill,inner sep=1.5pt]{};
			
			\draw [ultra thin,rotate=-23.5] (0,4) -- (0,0.4);
			\draw [ultra thin,rotate=-23.5] (0,-4) -- (0,-0.4);
			
			\draw[thick] (0,0) circle (0.4cm);
			\shade[ball color=green!60!blue,opacity=0.80] (0,0) circle (0.4cm);
			
			\draw[thick, black,<-,>=latex,rotate=-23.5] (0,0) [partial ellipse=360:180:0.68cm and 0.18cm];
			
			\foreach \angle in {0,...,360}
			\draw[black] [random dot=2*\angle:4cm];
			\draw[ultra thin] (4,0) arc(0:360:4cm and 4cm);
			
		\end{tikzpicture}
		\caption{Precession of the equinoctial points; vernal equinox at point V, as the primary direction, and the autumnal equinox at point A; geographic poles projected onto the celestial sphere as celestial poles.}\label{fig:pres}
	\end{figure}
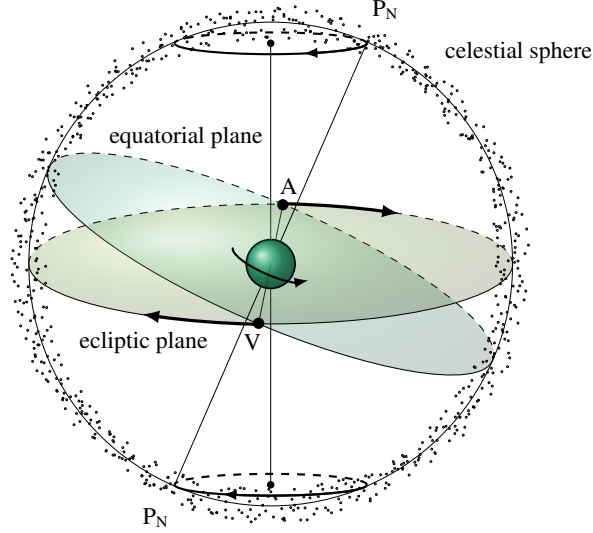

	The Moon contributes about $2/3$ of the total despite its small mass because the inverse-cube law favors proximity: $(r_\odot/r_{\mathrm{Moon}})^3 \approx (390)^3 \approx 5.9 \times 10^7$ overwhelms the mass ratio $m_\odot/m_{\mathrm{Moon}} \approx 2.7\times 10^7$ by a factor of $\sim 2.2$, so lunar torque is roughly $2.2$ times solar torque. Plugging in measured values gives the lunisolar precession in longitude
	\[
	\psi_A \approx 5039''/\mathrm{century}
	\]
	A separate effect of the slow motion of the ecliptic plane itself, driven by planetary perturbations on Earth's orbit reduces the equinox's net motion in longitude by $\sim10''$/century.\footnote{The honest decomposition is more involved than the arithmetic suggests: lunisolar precession is naturally defined along the equator; planetary precession acts on the ecliptic, and combining them into a single rate in ecliptic longitude requires a $\cos\epsilon$ projection plus small cross-terms. See \cite{Lieske1977} for the exact relationship between $\psi_A$, $\chi_A$, and $p_A$.} The result is general precession in longitude:
	\[
	p_A \approx 5028.8''/\mathrm{century} = 1.3969^\circ/\mathrm{century}
	\]
	The modern computational form is given in \cite{Lieske1977}; the calculation in full is given in \cite{BrouwerClemence1961}, including planetary contributions; Smart's treatment in \cite{Smart1977} is clean, and available for the lunisolar piece in isolation.
	
	\subsection{Connection to longitude polynomial}\label{subsec:longitudepolynomial}
	
	Three rates are now in play, shown in Table~\ref{tab:rates}.
	
	\begin{table}[h]
		\centering
		\caption{Rates of change for tropical, sidereal, and anomalistic mean motion}
		\label{tab:rates}
		\begin{tabular}{lll}
			\toprule
			\bfseries Rate & \bfseries Reference direction & \bfseries Value ($^\circ$/century) \\
			\midrule
			tropical mean motion ($L_0$ linear coefficient) & mean equinox of date & 36000.76983 \\
			sidereal mean motion & fixed stars & 35999.373 \\
			anomalistic mean motion ($M$ linear coefficient) & perihelion & 35999.05029 \\
			\bottomrule
		\end{tabular}
		
	\end{table}
	
	The differences are physical:
	\[
	36000.76983 - 35999.373 = 1.397^\circ/\mathrm{century} = 5028''/\mathrm{century} = p_A
	\]
	general precession in longitude, exactly as derived above. And
	\[
	35999.373 - 35999.05029 = 0.323^\circ/\mathrm{century} \approx 1162''/\mathrm{century} \approx 11.6''/\text{yr}
	\]
	the prograde precession of Earth's perihelion, also driven primarily by planetary perturbations (primarily Venus and Jupiter). The Meeus polynomial encodes three distinct astronomical phenomena in the difference between two linear coefficients. Note the reference frame: the $11.6''$/yr here is the apsidal advance against the fixed stars, period $360^\circ/11.6'' \approx 111{,}700$ yr. Relative to the precessing equinox the perihelion gains $50.29'' + 11.6'' = 61.9''$/yr and returns to a given solstice in $360^\circ/61.9'' \approx 21{,}000$ yr; it is this equinox-relative cycle, not the sidereal one, that governs the seasonal drift of $Y(L)$ in Section~\ref{sec:meanyears}.
	
	The quadratic term $0.0003032^\circ\,T^2$ in $L_0$ follows from acceleration of general precession itself: precession is not constant because the Earth--Moon--Sun geometry evolves on long timescales. Following \cite{Lieske1977,Capitaine2003}, general precession is
	\[
	p_A(T) = 5028.796''\,T + 1.105''\,T^2 + O(T^3)
	\]
	and the cumulative second-order contribution feeds the leading part of the $L_0$ quadratic.
	
	\subsection{Some remarks on precession}\label{subsec:remarksprecession}
	
	The natural way to feel the magnitude of precession is to ask: what is the difference between the tropical year (return to equinox-of-date) and the sidereal year (return to a fixed inertial direction)? It is exactly the time the Sun takes to traverse the precession increment, $50.29''$/yr at the mean motion $0.9856473^\circ$/day:
	\[
	P_{\mathrm{sid}} - P_{\mathrm{trop}} = \frac{50.29''/\text{yr}}{0.9856473^\circ/\mathrm{day}} \approx 0.01417 \mathrm{ day} \approx 20.4 \mathrm{ minutes/year}
	\]
	Over the J2000~$\pm$~1500~yr window, the cumulative offset between the two reference frames is $\sim 3000 \times 20.4$ min $\approx 42$ days, and the equinox direction has slid westward by $\sim 42^\circ$. The polynomial $L_0(T)$ tracks this automatically because it is referred to mean equinox of date. If we instead computed longitudes referred to mean equinox of J2000.0 and then asked when those returned to fixed values, we would be measuring a sidereal-style year ($\approx 365.2564$ d), not the tropical-family years we want (not to say that this is not valid, only that it is not the measurement we are interested in here).
	
	\section{Equation of center}\label{sec:eqncenter}
	
	The Sun's true anomaly $\nu$ is its actual angular position in its orbit, measured from perihelion, and differs from the mean anomaly $M$ because the Earth's orbit is elliptical. We need an explicit series for $\nu - M$ in powers of the eccentricity $e$ and harmonics of $M$.
	
	In this way, again, we follow from \cite{Smart1977}. Place the Sun at one focus of an ellipse with semi-major axis $a$, eccentricity $e$. Define:
	\begin{enumerate}
		\item \emph{mean anomaly} $M = n(t - t_p)$, where $n = 2\pi/P$ is the mean motion and $t_p$ is the time of perihelion passage; $M$ advances uniformly in time but does not point at the body, as it is a fictitious uniform-motion angle;
		\item \emph{eccentric anomaly} $E$, the angle measured from the ellipse's center to the body's position projected vertically onto the auxiliary circle (the circle of radius $a$ enclosing the ellipse);
		\item \emph{true anomaly} $\nu$, the actual angular position from the focus.
	\end{enumerate}
	
	Kepler's second law (equal areas in equal times), combined with the geometric definition of $E$, yields the celebrated \emph{Kepler equation} relating $M$ to $E$:
	\[
	M = E - e\sin E
	\]
	This is transcendental in $E$ and has no closed-form inverse. The geometric relation between $E$ and $\nu$, derived directly from the ellipse equation in polar form:
	\[
	\cos\nu = \frac{\cos E - e}{1 - e\cos E}, \qquad \tan\frac{\nu}{2} = \sqrt{\frac{1+e}{1-e}}\,\tan\frac{E}{2}
	\]
	
	We want $E$ as a function of $M$ and $e$. For Earth, $e \approx 0.0167$ is small,\footnote{$e$ is presently decreasing toward a minimum of $\sim 0.0023$ near $29{,}000$--$30{,}000$ CE \cite{laskar2004}; the $e(T)$ polynomial below carries the leading part of this decrease in its negative linear coefficient.} and we can develop $E$ as a power series in $e$. Lagrange's inversion theorem\footnote{From  \cite{weisstein_lagrange_inversion}: let $z$ be defined as a function of $w$ in terms of a parameter $\alpha$ by $z=w+ \alpha\varphi(z)$. Any function of $z$ can be expressed as a power series in $\alpha$ which converges for sufficiently small $\alpha$ and has the form:\[\begin{aligned}
			& F(z)=F(w)+\frac{\alpha}{1} \phi(w) F^{\prime}(w)+\frac{\alpha^2}{1 \cdot 2} \frac{\partial}{\partial w}\left\{[\phi(w)]^2 F^{\prime}(w)\right\} \\
			& +\ldots+\frac{\alpha^{n+1}}{(n+1)!} \frac{\partial^n}{\partial w^n}\left\{[\phi(w)]^{n+1} F^{\prime}(w)\right\}+\ldots
		\end{aligned}\].} applied to $E - M = e\sin E$ gives, to third order:
	\[
	E = M + e\sin M + \frac{e^2}{2}\sin 2M + \frac{e^3}{8}(3\sin 3M - \sin M) + O(e^4)
	\]
	The mechanical derivation is by repeated substitution: write $E^{(0)} = M$, $E^{(1)} = M + e\sin E^{(0)} = M + e\sin M$, $E^{(2)} = M + e\sin E^{(1)} = M + e\sin(M + e\sin M)$, expand $\sin(M + \delta) = \sin M\cos\delta + \cos M\sin\delta$, and collect to the desired order in $e$ \cite{BrouwerClemence1961}.
	
	Now we move from $E$ to $\nu$. Starting from $\tan(\nu/2) = \sqrt{(1+e)/(1-e)}\,\tan(E/2)$, expand the radical as
	\[
	\sqrt{\frac{1+e}{1-e}} = 1 + e + \frac{e^2}{2} + \frac{e^3}{2} + O(e^4)
	\]
	and use the half-angle relations together with the series for $E(M)$ above. After collecting in harmonics of $M$, the result is the equation of center:
	\[
	\nu - M = \left(2e - \frac{e^3}{4}\right)\sin M + \frac{5e^2}{4}\sin 2M + \frac{13e^3}{12}\sin 3M + O(e^4)
	\]
	To interpret the leading coefficient $2e$: at $M = 90^\circ$ (one quarter orbit past perihelion), the angular position of the body is ahead of the uniform-motion prediction by $2e$ radians. For Earth's $e = 0.016708634$, evaluating $(2e - e^3/4) \cdot 180/\pi$ gives $1.914602^\circ$, agreeing exactly to six decimal places with the leading coefficient in Meeus's polynomial below. The other two coefficients of the series likewise reproduce $5e^2/4 \cdot 180/\pi = 0.019995^\circ$ (Meeus: $0.019993$) and $13e^3/12 \cdot 180/\pi = 0.000290^\circ$ (Meeus: $0.000289$).
	
	Now substituting $e(T)$. The Earth's eccentricity is itself secularly varying, primarily under perturbations from Jupiter and Venus \cite{Meeus1998}:
	\[
	e(T) = 0.016708634 - 0.000042037\,T - 1.267{\times}10^{-7}\,T^2
	\]
	Substituting into the equation of center series and re-collecting in $T$ gives the form:
	\[
	C(T, M) = (1.914602 - 0.004817\,T - 0.000014\,T^2)\sin M + (0.019993 - 0.000101\,T)\sin 2M + 0.000289\sin 3M
	\]
	The Sun's true (geometric) longitude is then $\odot = L_0 + C$.
	
	\section{Apparent longitude: aberration and nutation}\label{sec:abberrationnutation}
	
	True longitude $\odot$ is referred to the mean equinox of date and is the \emph{geometric} direction to the Sun. What an observer sees is shifted by two further effects:
	\[
	\lambda = \odot - 0.00569^\circ - 0.00478^\circ\sin\Omega, \qquad \Omega(T) = 125.04^\circ - 1934.136^\circ\,T
	\]
	with $\Omega$ the longitude of the Moon's ascending node.
	
	\subsection{Aberration}\label{subsec:aberration}
	
	Light from a celestial body arrives at the observer along a direction shifted from the true line of sight by the observer's motion through the rest frame of the body.

	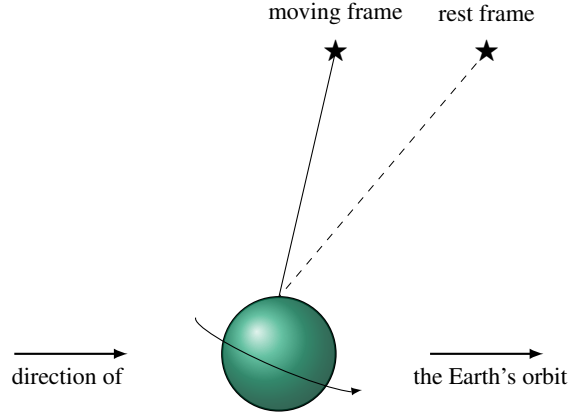
\begin{figure} [H]
		\centering
		\begin{tikzpicture}
			
			
			
			
			
			\node[] at (2.8,-0.3) {\small{the Earth's orbit}};
			\node[] at (-2.8,-0.3) {\small{direction of}};
			\node[] at (0.75,4) {$\bigstar$};
			\node[] at (2.75,4) {$\bigstar$};
			
			\draw [thin] (0.75,4) -- (0,0.75);
			\draw [ultra thin,dashed,] (2.75,4) -- (0,0.75);
			\node[] at (0.75,4.5) {\strut\small{moving frame}};
			\node[] at (2.75,4.5) {\small{rest frame}};
			
			\draw [thick,->,>=latex] (2,0) -- (3.5,0);
			\draw [thick,->,>=latex] (-3.5,0) -- (-2,0);
			
			\draw[thick] (0,0) circle (0.75cm);
			\shade[ball color=green!60!blue,opacity=0.80] (0,0) circle (0.75cm);
			
			\draw[black,<-,>=latex,rotate=-23.5] (0.0,0.0) [partial ellipse=360:180:1.2cm and 0.18cm];
			
		\end{tikzpicture}
		\caption{Apparent aberration; the apparent position of a star viewed from the Earth can change depending on the Earth's relative motion.}\label{fig:aberr}
	\end{figure}
	
	The classical result is that the apparent direction makes an angle $\theta'$ with the observer's velocity satisfying
	\[
	\tan\theta' = \frac{\sin\theta}{\cos\theta + v/c}
	\]
	where $\theta$ is the true angle. For Earth in orbit, $v_\oplus \approx 29.78$~km/s, and the Sun--Earth line is always perpendicular to the Earth's orbital velocity (the Earth moves tangentially around the Sun). The angle $\theta$ is, therefore, $90^\circ$ at all times, and the apparent Sun lags the true Sun in longitude (against the direction of the Earth's motion) by the constant of aberration \cite{Smart1977}:
	\[
	\kappa = \frac{v_\oplus}{c} = 20.4955'' \approx 0.005693^\circ
	\]
	Hence the constant $-0.00569^\circ$ correction. For an eccentric orbit Earth's orbital speed varies by $\pm e \approx \pm 1.7\%$ around the mean, modulating $\kappa$ by $\pm 0.3''$; Meeus drops this small variation in the low-precision form.
	
	\subsection{Nutation}\label{subsec:nutation}
	
	The same lunisolar torque on the Earth's equatorial bulge that drives steady precession has periodic components, because the Moon's orbital plane is not fixed: the Moon's ascending node $\Omega$ regresses around the ecliptic with period $18.6134$ years. As the geometry rotates, the magnitude and direction of the precessional torque oscillate, and the Earth's pole describes a small ellipse around the precessing mean position. The dominant period of this wobble is the Moon-node period.
	
	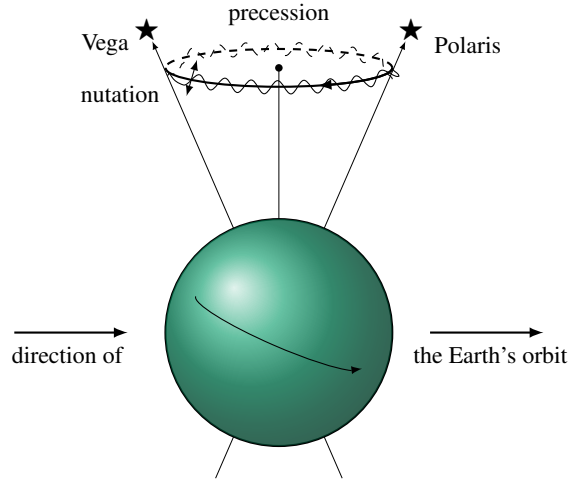
\begin{figure} [H]
		\centering
		\begin{tikzpicture}
			
			
			\draw [ultra thin] (0,3.5) -- (0,1.5);
			\node at (0,3.5)[circle,fill,inner sep=1.0pt]{};
			\draw[thick,black] (0,3.5) [partial ellipse=360:180:1.5cm and 0.25cm];
			\draw[thick,black,->,>=latex] (0,3.5) [partial ellipse=360:290:1.5cm and 0.25cm];
			\draw[thick,black,dashed] (0,3.5) [partial ellipse=0:180:1.5cm and 0.25cm];
			
			\draw[decoration={snake},decorate] (0,3.5) [partial ellipse=360:180:1.5cm and 0.25cm];
			\draw[dashed,decoration={snake},decorate] (0,3.35) [partial ellipse=0:180:1.5cm and 0.25cm];
			\draw[black, <->,>=latex, rotate=-11.5] (-1.78,3.32) -- (-1.82,2.82);
			
			\draw [ultra thin,<-,>=latex,rotate=-23.5] (0,4.2) -- (0,1.5);
			\draw [ultra thin,rotate=-23.5] (0,-2.1) -- (0,-1.5);
			\draw [ultra thin,<-,>=latex,rotate=23.5] (0,4.2) -- (0,1.5);
			\draw [ultra thin,rotate=23.5] (0,-2.1) -- (0,-1.5);
			
			\node[] at (0,4.2) {\small{precession}};
			\node[] at (2.5,3.8) {\small{Polaris}};
			\node[] at (-2.3,3.8) {\small{Vega}};
			\node[] at (2.8,-0.3) {\small{the Earth's orbit}};
			\node[] at (-2.8,-0.3) {\small{direction of}};
			\node[] at (1.75,4) {$\bigstar$};
			\node[] at (-1.75,4) {$\bigstar$};
			\node[] at (-2.1,3.15) {\small{nutation}};
			
			\draw [thick,->,>=latex] (2,0) -- (3.5,0);
			\draw [thick,->,>=latex] (-3.5,0) -- (-2,0);
			
			\draw[thick] (0,0) circle (1.5cm);
			\shade[ball color=green!60!blue,opacity=0.80] (0,0) circle (1.5cm);
			
			\draw[black,<-,>=latex,rotate=-23.5] (0.0,0.0) [partial ellipse=360:180:1.2cm and 0.18cm];
			
		\end{tikzpicture}
		\caption{Oscillatory motion of nutation superimposed on precession; nutation is in the direction of the plane defined by precession; changing of pole stars are shown to illustrate the effect of precession}\label{fig:nut}
	\end{figure}
	
	The nutation in longitude is the wobble of the equinox direction back and forth along the ecliptic, and is conventionally written as a sum of sinusoids in five fundamental angular arguments:
	
	\begin{enumerate}
		\item the Moon's mean anomaly;
		\item the Sun's mean anomaly;
		\item the Moon's argument of latitude;
		\item mean elongation Moon--Sun;
		\item node $\Omega$.
	\end{enumerate}
	
	The leading term, with amplitude exceeding all others by an order of magnitude, comes from the pure $\Omega$ argument \cite{Smart1977}:
	\[
	\Delta\psi \approx -17.20''\sin\Omega = -0.004778^\circ\sin\Omega
	\]
	This is the term Meeus retains \cite{Meeus1998}. The next-largest IAU 1980 nutation terms have amplitudes $\sim 1''$ (semiannual solar term in $2L_\odot$, monthly lunar term in $2L_{\mathrm{Moon}}$); the full IAU 1980 nutation series has 106 terms \cite{Seidelmann1982}. Their omission contributes $\lesssim 1''$ RMS to instantaneous $\lambda$, well within the stated $0.01^\circ$ precision. Collecting the pieces, the solar longitude used throughout is the Meeus low-precision model in closed form: the mean-longitude polynomial $L_0(T)$, the equation of center $C(M)$ truncated at $O(e^3)$ (three sine terms), the constant aberration $-0.00569^\circ$, and the single nutation term above. Note: it is not a multi-term periodic series; the count of 106 refers specifically to the IAU 1980 nutation expansion, of which we retain one term.

	\section{Root finding}\label{sec:rootfinding}
	
	For target longitude $L$, we want $t$ such that $\lambda(t) \equiv L \pmod{360^\circ}$. Define the wrapped residual
	\[
	\delta(t) = \big((\lambda(t) - L + 180^\circ) \bmod 360^\circ\big) - 180^\circ \in (-180^\circ, 180^\circ]
	\]
	and iterate
	\[
	t_{n+1} = t_n - \frac{\delta(t_n)}{\dot\lambda_\text{mean}}
	\]
	This is modified Newton: instead of the true Jacobian $\dot\lambda(t)$, which oscillates by $\pm 2e\approx \pm 3.3\%$ around the mean because of the equation of center, we use the constant mean motion. Convergence is, therefore, linear rather than quadratic (!), with contraction ratio
	\[
	\left|1 - \frac{\dot\lambda_\mathrm{true}(t)}{\dot\lambda_\mathrm{mean}}\right| \lesssim 2e \approx 0.033
	\]
	
	From an initial guess accurate to half a day, two or three iterations reach a $10^{-9}\,^\circ$ tolerance. \cite{ReingoldDershowitz2018} use the same fixed-point scheme in their \texttt{solar-longitude-after} function, also seeded by the mean motion.
	
	\section{Mean interval extraction and error analysis}\label{sec:meanerror}
	
	Let $t_1 < t_2 < \cdots < t_N$ be the successive crossings of longitude $L$ in the window. Define
	\[
	\bar Y_L = \frac{t_N - t_1}{N - 1}
	\]
	This is algebraically identical to the mean of consecutive intervals, $(N-1)^{-1}\sum_{k=1}^{N-1}(t_{k+1}-t_k)$, the intermediate $t_k$ telescope away (cancel pairwise).
	
	$\Delta T$ does not enter $\bar Y_L$. We work in Terrestrial Time, so each crossing $t_k$ is a TT instant and $\bar Y_L$ is a TT interval; $\Delta T = \mathrm{TT} - \mathrm{UT}$ never appears, and the telescoping above is an exact algebraic identity on the $t_k$, independent of the time scale. $\Delta T$ becomes relevant only if one re-expresses the result as a year in mean solar days, where the conversion is governed by the endpoint values of $\Delta T$ across the window. For J2000~$\pm$~1500~yr these are $\Delta T(500~\text{CE}) \approx 5.6 \times 10^3$~s and $\Delta T(3500~\text{CE}) \approx 9 \times 10^3$~s (\cite{Morrison2021}, amended from \cite{Morrison2016}; the future value is uncertain but of similar order), a difference of $\sim 3.4 \times 10^3$~s, hence $\sim 1$~s/yr over $N - 1 \approx 3000$ intervals. This residual does \emph{not} cancel, precisely because J2000 lies far from the vertex ($\sim 1820$) of the secular $\Delta T$ parabola; it is part of the length-of-day effect treated in Section~\ref{sec:secular}, and is exactly the contribution that working in TT removes from the present computation.
	
	Consider now why the $\sim 0.01^\circ$ formula error is also small. Treating the formula error in $\lambda$ as bounded oscillatory noise of amplitude $\sigma_\lambda \approx 0.01^\circ$, the induced error in each crossing time is $\sigma_t \approx \sigma_\lambda/\dot\lambda_\mathrm{mean} \approx 15$ minutes. By the same telescoping argument,
	\[
	\sigma_{\bar Y_L} \le \frac{2\sigma_t}{N-1} \approx \frac{30 \mathrm{ min}}{3000} \approx 0.6 \text{ s}
	\]
	The $\sim 2$~s residuals against Meeus's published cardinal-point values are within a small multiple of this bound. The residual follows from both the formula floor and the fact that Meeus's ``near year 2000'' tabulation is itself not at exactly the same epoch as our window midpoint. This argument is \emph{only} valid because $N$ is large; a single year length from one pair of consecutive crossings would carry the full $\sim 15$-minute uncertainty.
	
	\section{J2000 expansion}\label{sec:j2000b1900}
	
	The same Newcomb solar theory \cite{Newcomb1895} can be re-expanded around any epoch, yielding numerically distinct but mathematically equivalent polynomials. Older sources \cite{Meeus1988} and the first edition of \cite{Meeus1998} use B1900, JD 2415020.0:
	\[
	T_{1900} = \frac{\mathrm{JD} - 2415020.0}{36525}, \qquad L = 279.69668^\circ + 36000.76892^\circ\,T_{1900} + 0.0003025^\circ\,T_{1900}^2,\ \mathrm{etc.}
	\]
	The two expansions describe the same function $\lambda(\mathrm{JD})$ to within their respective truncation errors. Verifying consistency:
	
	\begin{table}[h]
		\centering
		\caption{Verification of consistency for B1900 and J2000}
		\label{tab:j2000b1900}
		\begin{tabular}{llll}
			\toprule
			\bfseries Quantity & \bfseries B1900 to +1 century & \bfseries J2000 & \bfseries Match \\
			\midrule
			$L$ secular advance & $0.76892^\circ$ & $0.76983^\circ$ & 2 decimal places \\
			$e$ at $T_{1900}=1$ & $0.01670924$ & $0.016708634$ & 5 places \\
			equation of center, $\sin M$ leading & $1.914671$ & $1.914602$ & 4 places \\
			aberration & $-0.00569^\circ$ & $-0.00569^\circ$ & exact \\
			nutation amplitude & $0.00479^\circ$ & $0.00478^\circ$ & 3 places \\
			\bottomrule
		\end{tabular}
		
	\end{table}
	
	For the lower rows ($e$, equation of center leading coefficient, nutation), residuals at the fourth or fifth decimal place are higher-order refit terms. The $L$ secular advance coefficients agree only to $2$ decimal places; the larger residual ($9 \times 10^{-4}$) follows from the accumulated effect of higher-order polynomial terms over the one-century extrapolation from B1900. Meeus retains a $T^3$ contribution in $M$ in the B1900 form (coefficient $-3.3 \times 10^{-6}$), which is dropped in the J2000 low-precision form because it contributes $\le 10^{-8}\,^\circ$ for $|T| \le 15$.
	
	Both forms are Taylor polynomials of an analytic function around their respective reference epochs. The polynomial of degree $d$ truncated around epoch $T_0$ has remainder $R_d(T) \sim f^{(d+1)}(\xi)(T - T_0)^{d+1}/(d+1)!$. The error grows as $|T - T_0|^{d+1}$. For our window (J2000 $\pm$ 1500 yr):
	
	\begin{itemize}
		\item J2000 form: $|T| \le 15$. The leading dropped term is $O(T^3)$ in $L_0$ with secular coefficient $\sim 10^{-8}\,^\circ$, giving a worst-case truncation error of order $10^{-8} \cdot 15^3 \approx 3 \times 10^{-5}\,^\circ$ in instantaneous longitude;
		
		\item B1900 form: $T_{1900} \in [-15, +16]$ for the \emph{same} dates, since B1900 is offset from J2000 by exactly one century; the truncation magnitude is essentially identical at the window edges, modulo a $\sim 20\%$ asymmetry $((16/15)^3 \approx 1.21)$.
	\end{itemize}
	
	The two forms are, therefore, not meaningfully different in worst-case error; they diverge in the \emph{shape} of the error:
	
	\begin{itemize}
		\item the J2000 expansion has zero error at $T = 0$ (the year 2000) and grows symmetrically outward; 
		
		\item the B1900 expansion has zero error at 1900 and is asymmetric across our window.
	\end{itemize}
	
	For interval averaging, the symmetric profile partially self-cancels in the endpoint difference $\varepsilon(t_N) - \varepsilon(t_1)$ from Section~\ref{sec:meanerror}, while the asymmetric profile does not.
	
	The difference in computed $\bar Y_L$ between the two forms is well under one second, quite invisible against the $\sim 2$s formula floor in Section~\ref{sec:meanerror}. We use the J2000 form for three reasons, in decreasing order of importance: (i) it matches the reference text \cite{Meeus1998}; (ii) the symmetric error profile gives slightly cleaner cancellation in the interval-averaging step; (iii) the mean-longitude epoch happens to land in the middle of our window of interest, so individual instantaneous longitudes are most accurate where the window is densest.
	
	\section{Recovery of eight mean year values}\label{sec:meanyears}
	
	Let us assemble the above material to calculate the mean year values at points along the ecliptic.
	
	\begin{enumerate}
		\item For each target longitude $L$ in the eight-element set, find the first crossing $t_1$ in the window using the Newton iterator of Section~\ref{sec:rootfinding}, seeded by an analytic estimate of the form $t_0 = T_\mathrm{start} + (L/360^\circ) \cdot 365.2422$~d.
		\item Estimate the number of intervals via $N - 1 \approx \lfloor(T_\mathrm{end} - t_1)/365.2422\rfloor$, and find the last crossing $t_N$ by Newton's method seeded at $t_1 + (N - 1) \cdot 365.2422$~d.
		\item Compute $\bar Y_L = (t_N - t_1)/(N-1)$.
	\end{enumerate}
	
	For the J2000~$\pm$~1500 yr window ($\sim$500 to 3500 CE), this yields:
	
	\begin{table}[h]
		\centering
		\caption{Derived values for the cardinal and cross-quarter days relative to Meeus}
		\label{tab:values}
		\begin{tabular}{lllll}
			\toprule
			$L$ & \bfseries Marker & \bfseries Type & \bfseries $\bar Y_L$ (d) & \bfseries $\Delta$ vs Meeus \\
			\midrule
			$0^\circ$   & March equinox     & cardinal      & $365.242364$ & $-0.49$~s \\
			$45^\circ$  & Beltane           & cross-quarter & $365.241932$ & --- \\
			$90^\circ$  & June solstice     & cardinal      & $365.241645$ & $+2.19$~s \\
			$135^\circ$ & Lughnasadh        & cross-quarter & $365.241685$ & --- \\
			$180^\circ$ & September equinox & cardinal      & $365.242025$ & $+1.29$~s \\
			$225^\circ$ & Samhain           & cross-quarter & $365.242453$ & --- \\
			$270^\circ$ & December solstice & cardinal      & $365.242722$ & $-1.56$~s \\
			$315^\circ$ & Imbolc            & cross-quarter & $365.242687$ & --- \\
			\bottomrule
		\end{tabular}
		
	\end{table}
	
	The cross-quarter dates are interpreted here as the ecliptic-longitude midpoints between the cardinal events, following the cleanest astronomical convention. Traditional Gaelic reckoning ties them to fixed calendar dates (1 February, 1 May, 1 August, 1 November) which drift against the longitudes; the two conventions agree to within a few days, but the longitude-based one is what the formalism naturally delivers.
	
	\subsection{Some remarks on mean year values}\label{subsec:remarksmeanyear}
	
	The four cardinal-point values agree with the corresponding intervals tabulated in \cite{Meeus2002} to within $\pm 2.2$s. The largest discrepancy (June solstice, $+2.19$~s) is consistent with the formula error bound established in Section~\ref{sec:meanerror}, and with the modest difference between Meeus's ``near year 2000'' tabulation epoch and our window midpoint.
	
	The eight-marker arithmetic mean is $365.24222$d, which overstates the mean tropical year by $\sim 3$s; with only eight samples, none near the true minimum at $L = 107^\circ$, the estimate is biased toward the high-$Y$ side of the curve. A more faithful estimate from the full $Y(L)$ sinusoid is discussed in Section~\ref{subsec:meanyeartable} below.
	
	\subsection{Longitude-averaged mean year}\label{subsec:meanyeartable}
	
	Since $Y(L)$ is near-sinusoidal, its longitude average is
	\[
	\langle Y \rangle = \frac{1}{2\pi}\int_0^{2\pi} Y(L)\,dL \;=\; P_0 + \frac{1}{2\pi}\int_0^{2\pi} A\cos(L - \varpi)\,dL + \cdots \;=\; P_0,
	\]
	the sinusoidal term integrating to zero. The longitude-averaged $\langle Y\rangle$, therefore, recovers the mean tropical year at epoch analytically; it is not equal to it by construction, but converges to it as the number of samples grows. Numerically, the Riemann sum over the $1^\circ$-resolution $Y(L)$ table (360 equally-spaced samples, J2000 $\pm$ 1500 yr) yields $\langle Y\rangle = 365.242189$d.
	
	Table~\ref{tab:meanyear} compares this result against three published reference values. The Laskar value \cite{Laskar1986} is the constant term of the series given in Section~\ref{subsec:changing}, evaluated at $T = 0$; Bretagnon and Rocher (2001) \cite{BretagnonRocher2001} give an independent series with a different constant term.
	
	\begin{table}[h]
		\centering
		\caption{Mean tropical year at J2000 from four sources: Laskar and Meeus--Savoie value is the constant term of the series in \cite{Laskar1986,MeeusSavoie1992}; Bretagnon and Rocher value is the constant term of their independent series \cite{BretagnonRocher2001}; Meeus value is from \cite{Meeus2002}. Offsets are (source $-$ $\langle Y\rangle$) $\times$ 86400.}
		\label{tab:meanyear}
		\begin{tabular}{llr}
			\toprule
			\bfseries Source & \bfseries Value (d) & \bfseries Offset from $\langle Y\rangle$ \\
			\midrule
			mean $Y(L)$       & $365.2421892$ & --- \\
			Laskar (1986), Meeus and Savoie (1992)  & $365.2421897$ & $+0.04$s \\
			Bretagnon and Rocher (2001)     & $365.2421905$ & $+0.11$s \\
			Meeus (2002)                   & $365.2421900$ & $+0.07$s \\
			\bottomrule
		\end{tabular}
		
	\end{table}
	
	All four values agree to within $0.12$s. The $\langle Y\rangle$ residual of $-0.04$ to $-0.11$s below the published series is systematic: the Riemann sum is converged to better than $0.001$s already at $1^\circ$ resolution. The offset follows from the small difference between the mean of $Y(L)$ as computed from the Meeus polynomial approximation and the constant terms of the published dynamical series, which are derived from independent fits to the Earth's orbital evolution.
	
	The value $Y(0^\circ) = 365.242364$d from Table~\ref{tab:values} deserves separate comment: it is the \emph{vernal-equinox year}, the mean interval between successive passages of the Sun through $L = 0^\circ$, and is \emph{not} the mean tropical year. The two differ at the present epoch. \cite{Meeus2002,HM2004} give a polynomial for the vernal-equinox year whose constant term at $T = 0$ is $365.2423748$d (uniform days), or $15.9$s longer than the Bretagnon and Rocher mean tropical year of \cite{BretagnonRocher2001}. Our computed $Y(0^\circ) = 365.242364$d lies $15.0$s above the Bretagnon and Rocher constant term and $0.9$~s below the Meeus vernal-equinox polynomial, consistent with the formula floor of $\sim 2$s from Section~\ref{sec:meanerror}. This is a non-trivial internal consistency check: the $Y(L)$ framework, applied at $L = 0^\circ$, independently recovers a value for the vernal-equinox year that agrees with the dedicated Meeus polynomial to within the stated precision of the computation.
	
	The physical origin is apsidal precession. Because the Earth's perihelion advances by $\sim 62''$/yr, the vernal equinox is encountered at a slightly later phase of the elliptical orbit each year; the Sun is moving slightly more slowly at that phase, prolonging the equinox-to-equinox interval relative to the longitude-averaged mean. The gap is not constant: Meeus's vernal-equinox polynomial has a positive linear coefficient ($+10.34 \times 10^{-5}$d/millennium), meaning the vernal-equinox year is currently \emph{increasing}, while the mean tropical year is decreasing (Section~\ref{subsec:changing}). Their difference is, therefore, epoch-dependent, and the $\sim 15$s figure applies specifically to J2000.
	
\begin{figure}[htbp]
		\centering
		\includegraphics[width=\textwidth]{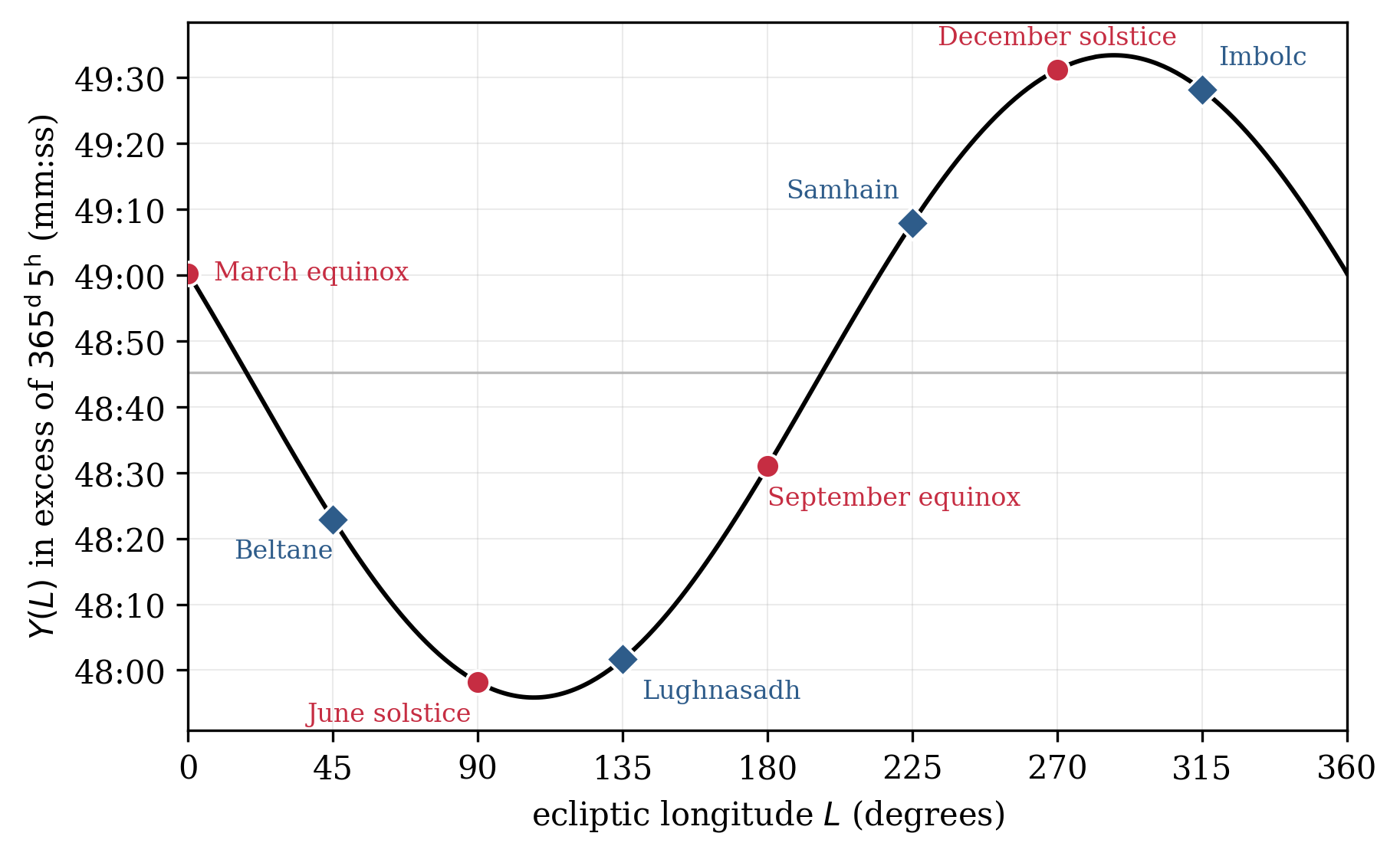}
		\caption{Mean year length $Y(L)$ as a function of ecliptic longitude $L$, computed at $1^{\circ}$ resolution over the window J2000 $\pm$ 1500~yr ($N \approx 3000$ intervals per longitude). The curve has minimum $Y = 365.241618$d at $L = 107^{\circ}$ and maximum $Y = 365.242748$d at $L = 288^{\circ}$, with peak-to-peak amplitude $97.56$s. These extrema lie $\sim17^{\circ}$ and $\sim8^{\circ}$ past the June and December solstices, respectively, because the Earth's perihelion currently lies at $L \approx 283^{\circ}$ (early January) and aphelion at $L \approx 103^{\circ}$; the extrema of $Y(L)$ track the perihelion and aphelion directions rather than the solstices. Markers show the eight sample values tabulated in Table~\ref{tab:values}: cardinal-point years in red dots; cross-quarter years in blue diamonds. Because the cardinal points at $L = 90^{\circ}$ and $270^{\circ}$ lie $\sim17^{\circ}$ from the curve's true extrema, the eight-marker amplitude ($\approx 93$s) understates the full peak-to-peak amplitude by $\approx4.6$s. The zero line in gray marks the mean tropical year: 365.24219d \cite{Laskar1986}.}
		\label{fig:ylambda}
	\end{figure}

	The full $Y(L)$ curve, computed at $1^{\circ}$ resolution and shown in Figure~\ref{fig:ylambda}, has its minimum at $L = 107^{\circ}$ ($Y = 365.241618$d) and its maximum at $L = 288^{\circ}$ ($Y = 365.242748$d), with peak-to-peak amplitude $97.56$s; these extrema are displaced $\sim17^{\circ}$ and $\sim5^{\circ}$ past the June and December solstices, respectively, and are not coincident with them. The eight cardinal and cross-quarter sample values in Table~\ref{tab:values} span only $93.0$s because the solstice markers at $L = 90^{\circ}$ and $270^{\circ}$ sample the curve $\sim17^{\circ}$ from its extrema; the full amplitude requires the continuous computation.
	
	The physical origin of both the pattern and the $17^{\circ}$ displacement lies in two slow drifts. The Earth's perihelion currently lies near $L \approx 283^{\circ}$ (early January, $\sim13^{\circ}$ past the December solstice) and its aphelion near $L \approx 103^{\circ}$ ($\sim17^{\circ}$ past the June solstice). The tropical year is shorter than the anomalistic year by approximately $1508$~s, so in each tropical year the Sun's mean anomaly falls short of a full $2\pi$ excursion by $\Delta \approx 3 \times 10^{-4}$rad. When a year begins at the perihelion longitude ($L \approx 283^{\circ}$), it ends with the mean anomaly $\Delta$ short of the next perihelion passage, i.e., in the decelerating phase of the equation of center, where the Sun's true longitude lags the mean. Reaching the original longitude, therefore, requires a small additional interval, lengthening $Y(L)$.  The symmetric argument applies at aphelion ($L \approx 103^{\circ}$): the year ends with the mean anomaly $\Delta$ short of the next aphelion, in the accelerating phase where the Sun slightly overshoots the mean longitude, shortening $Y(L)$. To leading order in eccentricity $e$: 
	
	\[ Y(L) \approx P_0 + \frac{\Delta \cdot 2e \cdot P_0}{2\pi} \cos\bigl(L - \varpi\bigr), \] 
	
	where $P_0$ is the mean tropical year, $e \approx 0.01671$, and $\varpi \approx 283^{\circ}$ is the perihelion longitude; this gives an amplitude of $\approx 101$s; the computed value of $97.56$s is $\sim 3\%$ smaller because the derivation holds $\varpi$ fixed, whereas over the $\pm 1500$yr window, the perihelion precesses by $\Delta\varpi \approx 26^\circ$ in each direction. Averaging $\cos(L - \varpi(t))$ over this drift reduces the effective amplitude by the factor $\sin(\Delta\varpi)/\Delta\varpi \approx 0.967$, recovering $101 \times 0.967 \approx 97.4$s, within $0.2$s of the computed value.
	
	The extrema of $Y(L)$ track perihelion and aphelion, not the points $90^{\circ}$ away from them. The $\sim17^{\circ}$ displacement of the extrema from the solstices will increase by $\sim 1.7^{\circ}$ per century as perihelion precesses; over $\sim 5200$ years, the extrema will have migrated to near the equinox longitudes!
	
	The near-alignment of perihelion with the December-solstice direction, and, hence, the proximity of the year-length extremum to $L=270^{\circ}$, is a feature of the present epoch; the perihelion-to-equinox drift of $61.9''$/yr accumulates to $\sim 25.8^\circ$ per half-window, or $\sim 51.6^\circ$ across the full 3000-year span; the window-averaged $Y(L)$, therefore, integrates over a substantial range of perihelion positions. The table's row-by-row labels remain valid as fixed ecliptic-longitude designations throughout; what changes within the window is which longitude carries the extremal year length. Over $\sim 5200$ years, the drift accumulates to a full $90^\circ$, at which point the extrema migrate to the equinox longitudes and the cross-quarter values become the new locations of cardinal-marker year lengths. Any narrative tying $\bar Y_L$ extrema to specific calendar markers, therefore, has an implicit epoch attached; the numerical recovery itself is epoch-independent at fixed $L$.
	
	\subsection{Epoch dependence of marker year lengths}\label{subsec:epochtable}
	
\begin{figure}[htbp]
		\centering
		\includegraphics[width=\textwidth]{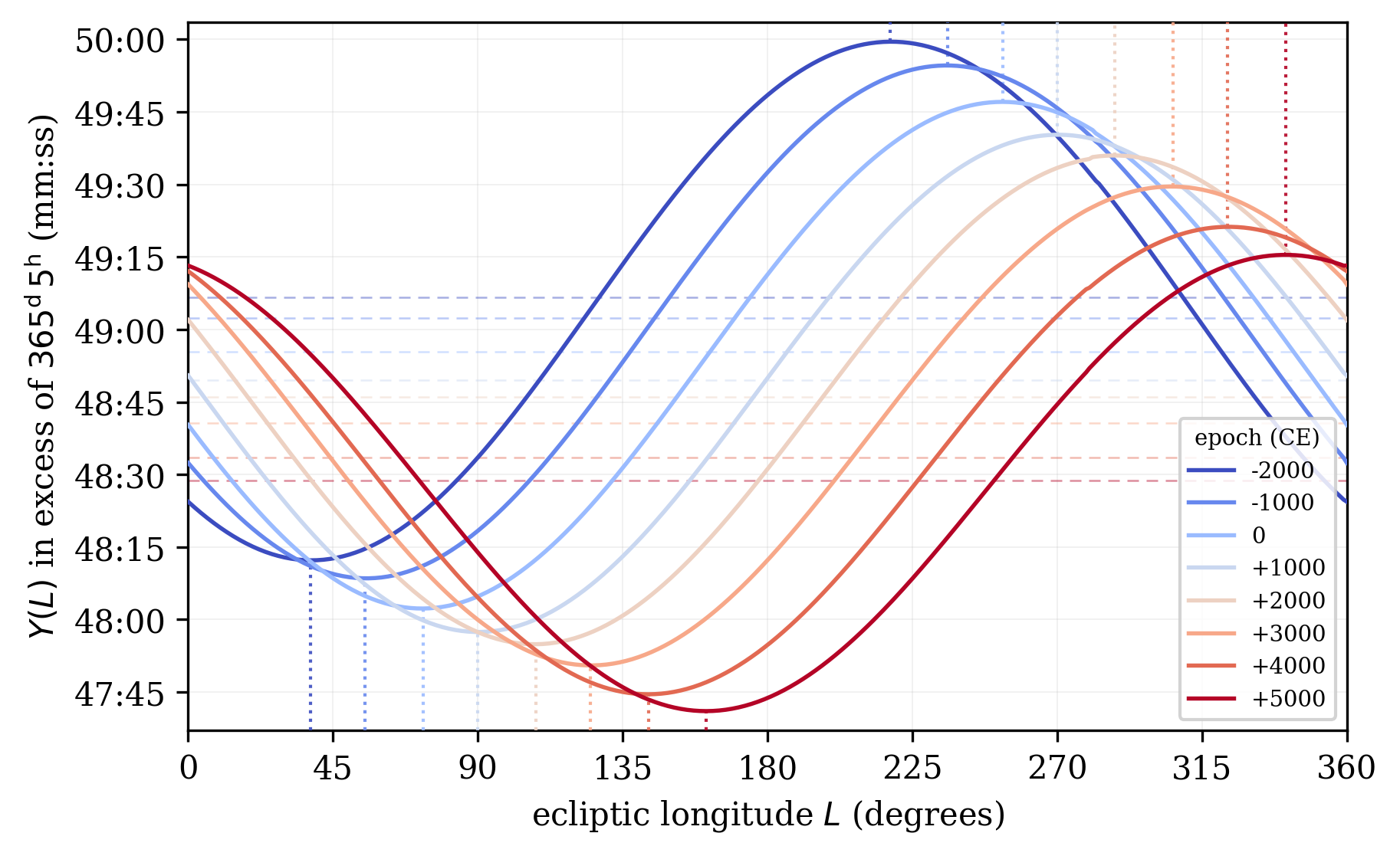}
		\caption{$Y(L)$ sinusoids at six center epochs ($-2000$ to $+5000$ CE, half-window $\pm 250$ yr each), computed at $1^\circ$ resolution. Colors run from blue (epoch $-2000$) to red (epoch $+5000$) via a diverging colormap. Dotted verticals mark the extrema of each curve; their progression from left to right traces apsidal precession. The $Y$-axis is offset from the mean tropical year $P_0 = 365.24219$d \cite{Laskar1986}; note that the overall level of each curve also drifts slightly as the mean tropical year shortens (Section~\ref{subsec:changing}).}
		\label{fig:epochoverlay}
	\end{figure}

	Since the $Y(L)$ sinusoid shifts rigidly in phase as perihelion precesses, each marker's mean year length changes substantially over historical and future timescales. Table~\ref{tab:epochmarkers} shows $Y(L)$ at all eight markers for eight center epochs spanning $-2000$ to $+5000$ CE, each computed with a $\pm 250$yr averaging window (analogous to \cite{HM2004} Table~4, here extended to cross-quarter days). Figure~\ref{fig:epochoverlay} overlays the corresponding $Y(L)$ sinusoids; the dotted verticals mark each curve's minimum, making the phase migration directly visible.
	
	\begin{table}[h]
		\centering
		\caption{$Y(L)$ (days) at all eight ecliptic markers by center epoch, half-window $\pm 250$ yr. Values computed with the Newton-iteration procedure of Section~\ref{sec:rootfinding}. Epoch $0$ row matches Table~\ref{tab:values} to within the formula floor of $\sim 2$~s. Compare \cite{HM2004} Table~4 for the four cardinal points.}
		\label{tab:epochmarkers}
		\resizebox{\textwidth}{!}{%
			\begin{tabular}{r cccc cccc}
				\toprule
				\bfseries Epoch & \bfseries March equinox & \bfseries Beltane & \bfseries June solstice & \bfseries Lughnasadh
				& \bfseries September equinox   & \bfseries Samhain & \bfseries December solstice  & \bfseries Imbolc \\
				\bfseries (CE)  & $L=0^\circ$        & $L=45^\circ$      & $L=90^\circ$       & $L=135^\circ$
				& $L=180^\circ$      & $L=225^\circ$     & $L=270^\circ$      & $L=315^\circ$ \\
				\midrule
				$-2000$ & $365.241949$ & $365.241813$ & $365.242055$ & $365.242517$ & $365.242922$ & $365.243046$ & $365.242824$ & $365.242373$ \\
				$-1000$ & $365.242042$ & $365.241776$ & $365.241879$ & $365.242283$ & $365.242737$ & $365.242983$ & $365.242891$ & $365.242509$ \\
				$\phantom{+}0$    & $365.242133$ & $365.241766$ & $365.241722$ & $365.242030$ & $365.242493$ & $365.242838$ & $365.242881$ & $365.242593$ \\
				$+1000$ & $365.242251$ & $365.241820$ & $365.241637$ & $365.241818$ & $365.242245$ & $365.242658$ & $365.242827$ & $365.242661$ \\
				$+2000$ & $365.242384$ & $365.241936$ & $365.241636$ & $365.241676$ & $365.242027$ & $365.242469$ & $365.242748$ & $365.242716$ \\
				$+3000$ & $365.242469$ & $365.242046$ & $365.241666$ & $365.241566$ & $365.241809$ & $365.242240$ & $365.242602$ & $365.242696$ \\
				$+4000$ & $365.242501$ & $365.242139$ & $365.241719$ & $365.241493$ & $365.241606$ & $365.241983$ & $365.242393$ & $365.242601$ \\
				$+5000$ & $365.242514$ & $365.242245$ & $365.241827$ & $365.241505$ & $365.241479$ & $365.241765$ & $365.242182$ & $365.242487$ \\
				\bottomrule
		\end{tabular}}
	\end{table}
	
	Several qualitative features in Table~\ref{tab:epochmarkers} are immediately readable from the phase-migration picture:
	
	\begin{itemize}
		\item the March equinox year $Y(0^\circ)$ increases monotonically by $\sim 49$s over the 7000-year span, as the $Y(L)$ maximum migrates toward $L = 0^\circ$ from higher longitudes; this is the physically correct expectation: the vernal-equinox year is lengthening because perihelion is approaching that longitude;
		
		\item the September equinox year $Y(180^\circ)$ decreases monotonically by $\sim 124$s, because the $Y(L)$ minimum is migrating through that longitude; by epoch $+5000$, the September equinox carries the shortest year of the eight markers, a role held by the June solstice at epoch $-2000$;
		
		\item the June solstice year $Y(90^\circ)$ is non-monotonic: it first falls as the minimum approaches $L = 90^\circ$ (reaching its lowest sampled value near epoch $+1000$--$+2000$) and then rises again as the minimum migrates past; the cross-quarter markers show analogous non-monotonic behavior offset by $45^\circ$;
		
		\item the maximum of $Y(L)$ at epoch $-2000$ falls near Samhain ($L = 225^\circ$, value $365.243046$~d); by epoch $+5000$ the maximum has migrated to Imbolc ($L = 315^\circ$) and beyond, while Samhain has become one of the shorter-year markers; the $\sim 130^\circ$ of phase migration over 7000 years is consistent with $1.7^\circ/\mathrm{century} \times 70$ centuries.
	\end{itemize}
	
	The epoch dependence has a direct calendrical implication in the following way. A calendar designed to minimize drift against a specific marker, say, the March equinox, should be evaluated against the $Y(L)$ value at that marker for its reform epoch, not the J2000 baseline value. The difference between the two can reach $\sim 15$s for a reform epoch a millennium away from J2000, corresponding to a $\sim 15$-second shift in the effective ``zero-drift'' mean year for that marker. Leap-rule analyses that use a single epoch-independent $Y(L)$ value absorb this shift silently into the residual drift.
	
	\section{A note on secular drift and quadratic limit}\label{sec:secular}
	
	We conclude with a remark on secular drift and the quadratic limit it imposes on all fixed-rule calendars. For the sake of calendrical systems, a constant mean year $\bar{Y}$ is chosen; the true tropical year is not constant. We derive the drift equation and show why no fixed-rule calendar can escape the resulting quadratic error.
	
	\subsection{Changing tropical year}\label{subsec:changing}
	
	The mean tropical year $Y(T)$ as a function of time was computed by \cite{Laskar1986} from the long-term evolution of the Earth's orbital elements under gravitational perturbations. The full series \cite{MeeusSavoie1992} is:
	\[
	Y(T) = 365.2421896698 - 6.15359 \times 10^{-6}\,T - 7.29 \times 10^{-10}\,T^2 + 2.64 \times 10^{-10}\,T^3,
	\]
	where $T$ is Julian centuries from J2000.0. The physical driver is the secular change in the rate of equinoctial precession. Over the few-millennia span relevant here, the quadratic and cubic terms are small relative to the linear term, giving the truncation
	\[
	Y(T) \;\approx\; Y_0 - r\,T, \qquad Y_0 = 365.242189, \quad r \approx 6.16 \times 10^{-6} \mathrm{ days/century} \approx 0.532 \mathrm{ s/century}.
	\]
	Among the higher-order terms the cubic dominates: its coefficient is only $\sim$3 times smaller, so it overtakes the quadratic by $\sim$300 yr. What sets the cumulative drift below is instead the size relative to the \emph{linear} term $r$, and the cubic reaches that magnitude only near $\sim$15{,}000 yr; the linear truncation of $Y(T)$ is, therefore, accurate to a few percent over the few-millennia floor, with residual quantified in Section~\ref{subsec:floor}. The cubic coefficient is positive: the shrinkage decelerates and, eventually, reverses, the basis of the range note in Section~\ref{subsec:floor}. Beyond $\sim$15{,}000 yr, the polynomial fit is unreliable regardless. The tropical year is \emph{shrinking} (!): it loses about half a second per century; see \cite{ReingoldDershowitz2018} for a figure representation of this change.
	
	\cite{BretagnonRocher2001,HM2004} give an independent series in Julian millennia from J2000.0:
	\[
	Y(T_m) = 365.24219052 - 6.156 \times 10^{-5}\,T_m - 6.84 \times 10^{-8}\,T_m^2 + 2.630 \times 10^{-7}\,T_m^3 + 3.2 \times 10^{-9}\,T_m^4,
	\]
	where $T_m$ is Julian millennia from J2000.0. Converting the linear coefficient to centuries ($-6.156 \times 10^{-5}$d/millennium $= -6.156 \times 10^{-6}$d/century) gives agreement with the Laskar and Meeus and Savoie rate to three significant figures. The constant terms differ by $365.24219052 - 365.2421896698 = +0.0000009$d $\approx +0.073$s: a genuine discrepancy between two independent fits to the Earth's orbital evolution. Both series appear in Table~\ref{tab:meanyear}.
	
	\subsection{Drift ODE}\label{subsec:ode}
	
	A fixed-rule calendar has constant mean year $\bar{Y}$. The true tropical year at century $T$ is $Y(T) = Y_0 - rT$. Each year the seasonal marker slips by the \emph{per-year} difference $\bar{Y} - Y(T)$ days; a Julian century contains $100$ years, so the cumulative drift accrues at the per-century rate
	\[
	\frac{d\Delta}{dT} = 100\,\bigl[\bar{Y} - Y(T)\bigr] = 100\,\bigl[\underbrace{(\bar{Y} - Y_0)}_{\varepsilon_0} + r\,T\bigr].
	\]
	The factor of $100$ is essential: $\bar{Y} - Y(T)$ is a length difference \emph{per year}, whereas $T$ is measured in centuries. This is the governing ordinary differential equation. The bracket has two terms:
	\begin{itemize}
		\item the initial mismatch $\varepsilon_0$ between the calendar's mean year and the tropical year at epoch; this is constant, in that it depends only on how well a leap rule is tuned;
		
		\item the secular term $r\,T$, growing linearly with time; even if $\varepsilon_0 = 0$ (a perfectly tuned calendar), this term ensures that the drift rate increases without bound.
		
	\end{itemize}
	
	Integrating the ODE from $T = 0$:
	\[
	\Delta(T) = 100\int_0^T \bigl[\varepsilon_0 + r\,t\bigr]\,dt = 100\,\varepsilon_0\,T + 50\,r\,T^2.
	\]
	
	Two contributions to the cumulative drift:
	\begin{enumerate}
		\item linear term $100\,\varepsilon_0\,T$ from the constant mismatch is what leap rules are designed to minimize; a better leap rule makes $|\varepsilon_0|$ smaller, reducing this term;
		\item quadratic term $50\,r\,T^2$ from the changing tropical year is \emph{independent of the leap rule}, in that no choice of $\bar{Y}$ can eliminate it; it grows as the square of time, and eventually dominates.
	\end{enumerate}
	
	
	Differentiating the ODE once more:
	\[
	\frac{d^2\Delta}{dT^2} = 100\,r \approx 6.16 \times 10^{-4} \mathrm{ days/century}^2.
	\]
	The drift acceleration is constant. The error rate grows linearly; the cumulative error grows quadratically. This is the reason that no fixed rule can track a moving target: the calendar produces a constant $\bar{Y}$ (zeroth-order), which can cancel a constant drift rate (first-order), but cannot cancel an accelerating drift (second-order).
	
	\subsection{Secular floor (orbital and length-of-day)}\label{subsec:floor}
	
	The quantity that shrinks at rate $r$ is the tropical year measured as a \emph{Terrestrial Time} (SI-second) interval, an orbital quantity. Setting $\Delta = 1$~day and keeping only the quadratic term,
	\[
	50\,r\,T^{2} = 1 \quad\Longrightarrow\quad
	T = \sqrt{\frac{1}{50\,r}}
	= \sqrt{\frac{1}{50 \cdot 6.16 \times 10^{-6}}}
	\approx 57 \text{ centuries} = 5{,}700 \text{ years}.
	\]
	This $5{,}700$-year figure is, therefore, the \emph{orbital} floor: it is what a fixed-rule calendar would suffer if civil days were of fixed SI length.
	
	They are not; a civil calendar counts mean solar days, and the mean solar day is lengthening through tidal friction at $\dot D \approx 1.8$~ms/century \cite{Morrison2021}. Writing the year in mean solar days as $Y_{\mathrm{d}} = Y_\tau / D$ ($Y_\tau$ the year in TT seconds, $D$ the mean solar day), its secular rate carries a second term,
	\[
	\dot Y_{\mathrm{d}}
	= \frac{\dot Y_\tau}{D} - \frac{Y_\tau}{D^{2}}\dot D
	= -\,r \;-\; \frac{Y_{\mathrm{d}}}{D}\,\dot D,
	\qquad
	\frac{Y_{\mathrm{d}}}{D}\,\dot D \approx 7.6 \times 10^{-6}\ \mathrm{d/century},
	\]
	comparable to, and somewhat larger than, the orbital $r = 6.16 \times 10^{-6}$. The combined rate $r' \approx 1.4 \times 10^{-5}$~d/century is roughly twice $r$, and the corresponding floor is
	\[
	\sqrt{\frac{1}{50\,r'}} \approx 38 \text{ centuries} \approx 3{,}800 \text{ years}.
	\]
	The present work computes everything in TT at the expense of the length-of-day term throughout (Section~\ref{sec:meanerror}); this is harmless for the \emph{instantaneous} year lengths and for $\bar Y_L$, where $\Delta T$ telescopes out of the endpoint difference, but it is decidedly \emph{not} harmless for this secular conclusion, since working in TT removes precisely the length-of-day contribution that a civil calendar feels. The floor for a day-counting calendar is, therefore, the shorter $\sim 3{,}800$-year figure, with the orbital $5{,}700$-year value as its uniform-time limit.
	
	Both floors keep only the leading quadratic-in-$T$ drift term. Carrying the quadratic and cubic terms of $Y(T)$ through the same integral $\Delta = 100\int(\bar Y - Y)\,dt$ adds
	\[
	\Delta_{\mathrm{ho}}(T) = -\tfrac{100}{3}\,a_2\,T^3 - 25\,a_3\,T^4,
	\qquad a_2 = -7.29\times10^{-10},\quad a_3 = +2.64\times10^{-10},
	\]
	the coefficients of $T^2$ and $T^3$ in $Y(T)$. At the orbital floor ($T \approx 57$) these net to $\sim -0.07$~d, the $T^4$ term dominating, lengthening it to $\sim 5{,}900$~yr ($+4\%$); at the shorter civil floor ($T \approx 38$) the same terms contribute only $\sim -0.014$~d ($+0.6\%$), since $T^4$ grows steeply with $T$. The correction is governed by the positive cubic coefficient $a_3$, the same term that reverses the shrinkage below; we quote the leading-order $5{,}700$ and $3{,}800$~yr as accurate to a few percent.
	
	A fixed intercalation rule yields a rational constant mean year $\bar Y = 365 + \sum_i a_i/n_i$; it can null the constant mismatch $\varepsilon_0$ by choosing $\bar Y \approx Y_0$, canceling the linear drift $100\,\varepsilon_0 T$, but it cannot cancel the accelerating term $50\,r'T^{2}$, which demands a time-varying mean year. Whether that variation is worth introducing, and by what mechanism (a scheduled change of leap rule, an observational or astronomical rule, or a progressive arithmetic rule), is a question the present geometric treatment does not settle. A zeroth-order constant cancels a first-order drift, but not a second-order one, regardless of how the leap rule is tuned.
	
	Finally, a note on range. The quadratic model assumes $r$ constant, but $r$ is itself slowly varying and ultimately changes sign as eccentricity, obliquity, and the precession rate evolve (the positive cubic coefficient $a_3$ above already implies $dY/dT = 0$ near $T \approx 89$ centuries); long-term integrations place the reversal of the mean-tropical-year trend on the order of $10^4$ yr hence. The floor derived above lies well inside the regime where $r$ is effectively constant, but the linear-in-$T$ model of $Y(T)$, and hence this extrapolation, must not be pushed past $\sim 10^4$ yr.
	
	\newpage
	\bibliography{references}

@book{Meeus1998,
	author    = {Meeus, Jean},
	title     = {Astronomical Algorithms},
	edition   = {2nd},
	publisher = {Willmann-Bell},
	year      = {1998}
}

@book{Meeus2002,
	author    = {Meeus, Jean},
	title     = {More Mathematical Morsels},
	publisher = {Willmann-Bell},
	year      = {2002}
}

@book{Meeus1988,
	author    = {Meeus, Jean},
	title     = {Astronomical Formulae for Calculators},
	edition   = {4th},
	publisher = {Willmann-Bell},
	year      = {1988}
}

@book{Smart1977,
	author    = {Smart, W. M.},
	title     = {Textbook on Spherical Astronomy},
	edition   = {6th},
	note      = {Revised by R. M. Green},
	publisher = {Cambridge University Press},
	year      = {1977}
}

@book{BrouwerClemence1961,
	author    = {Brouwer, Dirk and Clemence, Gerald M.},
	title     = {Methods of Celestial Mechanics},
	publisher = {Academic Press},
	year      = {1961}
}

@article{Lieske1977,
	author  = {Lieske, J. H. and Lederle, T. and Fricke, W. and Morando, B.},
	title   = {Expressions for the precession quantities based upon the {IAU} (1976) system of astronomical constants},
	journal = {Astronomy \& Astrophysics},
	volume  = {58},
	year    = {1977},
	pages   = {1--16}
}

@book{ReingoldDershowitz2018,
	author    = {Reingold, Edward M. and Dershowitz, Nachum},
	title     = {Calendrical Calculations: The Ultimate Edition},
	edition   = {4th},
	publisher = {Cambridge University Press},
	year      = {2018}
}

@article{Newcomb1895,
	author  = {Newcomb, Simon},
	title   = {Tables of the Motion of the Earth on Its Axis and Around the Sun},
	journal = {Astronomical Papers of the American Ephemeris},
	volume  = {VI},
	number  = {I},
	year    = {1895}
}

@ARTICLE{Capitaine2003,
	author = {Capitaine, N. and Wallace, P.~T. and Chapront, J.},
	title = {Expressions for IAU 2000 precession quantities},
	journal = {Astronomy and Astrophysics},
	year = {2003},
	volume = {412},
	pages = {567--586},
	doi = {10.1051/0004-6361:20031539},
}

@article{Morrison2021,
	author = {Morrison, L. V. and Stephenson, F. R. and Hohenkerk, C. Y. and Zawilski, M.},
	title = {Addendum 2020 to `Measurement of the Earth’s rotation: 720 BC to AD 2015'},
	journal = {Proceedings of the Royal Society A: Mathematical, Physical and Engineering Sciences},
	volume = {477},
	number = {2246},
	pages = {20200776},
	year = {2021},
	month = {02},
	issn = {1364-5021},
	doi = {10.1098/rspa.2020.0776},
}

@article{Morrison2016,
	author = {Stephenson, F. R. and Morrison, L. V. and Hohenkerk, C. Y.},
	title = {Measurement of the Earth's rotation: 720 BC to AD
	2015},
	journal = {Proceedings of the Royal Society A: Mathematical, Physical and Engineering Sciences},
	volume = {472},
	number = {2196},
	pages = {20160404},
	year = {2016},
	month = {12},
	issn = {1364-5021},
	doi = {10.1098/rspa.2016.0404},
}

@article{Laskar1986,
	title={Secular terms of classical planetary theories using the results of general theory},
	author={Jacques Laskar},
	journal={Astronomy and Astrophysics},
	year={1986},
	volume={157},
	pages={59--70},
}

@article{laskar2004,
	author  = {Laskar, J. and Robutel, P. and Joutel, F. and Gastineau, M. and Correia, A. C. M. and Levrard, B.},
	title   = {A long-term numerical solution for the insolation quantities of the Earth},
	journal = {A\&A},
	volume  = {428},
	number  = {1},
	pages   = {261--285},
	year    = {2004},
	doi     = {10.1051/0004-6361:20041335}
}

@ARTICLE{MeeusSavoie1992,
	author = {Meeus, Jean and Savoie, Denis},
	title = {The history of the tropical year},
	journal = {Journal of the British Astronomical Association},
	year = {1992},
	volume = {102},
	pages = {40--42},
}

@book{Hutton2001,
	author    = {Hutton, Ronald},
	title     = {The Triumph of the Moon: A History of Modern Pagan Witchcraft},
	publisher = {Oxford University Press},
	year      = {2001},
	isbn      = {978-0192854490}
}

@article{McCluskey1989,
	author = {Stephen C. McCluskey},
	title ={The Mid-Quarter Days and the Historical Survival of {B}ritish Folk Astronomy},
	journal = {Journal for the History of Astronomy},
	volume = {20},
	number = {13},
	pages = {S1--S19},
	year = {1989},
	doi = {10.1177/002182868902001302},
}

@misc{weisstein_lagrange_inversion,
	author       = {Weisstein, Eric W.},
	title        = {Lagrange Inversion Theorem},
	howpublished = {MathWorld--A Wolfram Resource},
	url          = {https://mathworld.wolfram.com/LagrangeInversionTheorem.html}
}

@article{BretagnonRocher2001,
	author  = {Bretagnon, Pierre and Rocher, Patrick},
	title   = {Du Temps universel au Temps coordonn{\'e}e barycentrique},
	journal = {Revue du Palais de la D{\'e}couverte},
	volume  = {285},
	year    = {2001},
	month   = {f{\'e}vrier},
	pages   = {39}
}

@misc{HM2004,
	title={A concise review of the {I}ranian calendar}, 
	author={M. Heydari-Malayeri},
	year={2004},
	eprint={astro-ph/0409620},
	archivePrefix={arXiv},
	primaryClass={astro-ph},
	url={https://arxiv.org/abs/astro-ph/0409620}, 
}

@article{Seidelmann1982,
	author  = {Seidelmann, P. K.},
	title   = {1980 {IAU} Theory of Nutation: The Final Report of the {IAU} Working Group on Nutation},
	journal = {Celestial Mechanics},
	volume  = {27},
	number  = {1},
	pages   = {79--106},
	year    = {1982},
	doi     = {10.1007/BF01228952}
}

\end{document}